\documentclass[final,amssymb,amsmath]{aipproc}
\usepackage{graphicx}
\usepackage[rflt]{floatflt}
\usepackage[vcentermath]{youngtab}
\usepackage{mathptm}
\layoutstyle{6x9}

\renewcommand{\theequation}{\arabic{equation}}

\newcommand{\pvalt}{\raise0.15ex\hbox{-}\mkern-11.5mu\int}
\newcommand{\be}{\begin{equation}}
\newcommand{\ee}{\end{equation}}
\newcommand{\bea}{\begin{eqnarray}}
\newcommand{\eea}{\end{eqnarray}}
\newcommand{\ben}{\begin{enumerate}}
\newcommand{\een}{\end{enumerate}}
\newcommand{\bit}{\begin{itemize}}
\newcommand{\eit}{\end{itemize}}

\newcommand{\half}{\frac{1}{2}}

\renewcommand{\ln}{\,\mbox{ln}\,}

\newcommand{\beq}{\begin{equation}}
\newcommand{\eeq}{\end{equation}}
\newcommand{\ba}{\begin{array}}
\newcommand{\ea}{\end{array}}

\newcommand{\<}{\langle}
\renewcommand{\>}{\rangle} 

\begin{document}


\title{Color Non-Singlet Spectroscopy}
 
\author{R.~L.~Jaffe}{
address={Center for Theoretical Physics,\\
Laboratory for Nuclear Science and Department of Physics\\
Massachusetts Institute of Technology, \\Cambridge, Massachusetts 02139},
}
\rightline{MIT-CTP-3663}

\begin{abstract} 
Study of the spectrum and structure of color non-singlet combinations of quarks and antiquarks, neutralized by a non-dynamical compensating color source, may provide an interesting way to address questions about QCD that cannot be addressed by experiment at the present time.  These states can be simulated in lattice QCD and the results can be used to improve phenomenological models of hadrons.  Here these ideas are applied to color triplet states of $qqqq$ and $qq\bar q$.
\end{abstract}

\maketitle 
 
\section{Introduction} 
\label{section0} 

All hadrons are, of course, color singlets.  However, interesting questions about quark correlations in QCD may be studied by considering what amounts to  the spectrum of non-singlet configurations of quarks and antiquarks\footnote{Hybrid states including gluons are also of interest, but will not be considered further here.}, neutralized by a non-dynamical compensating color source.  This spectrum can be studied on the lattice and with phenomenological models.  The two approaches can be compared with each other, and, in principle, with $B$ mesons and baryons, if a concerted effort were made to develop this area of spectroscopy.
 
Interest in unusual states of quarks has been piqued by recent reports of exotic baryons\cite{theta,reviews} and unexpected mesons\cite{charmstates}.  Although some of these reports are controversial and may prove incorrect, they have raised important questions about the quark substructure of matter.  Quark correlations, in particular, play an important role in the discussion.  Long ago it was recognized that the spin dependent one-gluon exchange force in QCD, known as ``colormagnetism'', pairs quarks in the antisymmetric flavor configuration\cite{rjmulti}.  More recently, reports of the exotic $\Theta^{+}$ baryon motivated a ``schematic'' diquark model\cite{jw,sn}, in which two quarks correlated in the color, flavor, and spin antisymmetric channel --- the ``good'' diquark of Ref.~\cite{jw} --- was assumed to dominate the light hadron spectrum.   Although similar, these two are not the same:  the flavor antisymmetric diquark can include both color sextet (spin one) and color antitriplet (spin zero) components, while the ``good'' diquark is assumed to be pure color antitriplet.  Other correlations have been proposed.  For example Karliner and Lipkin have suggested that ``triquark'' correlations ($qq\bar q$) are very important in spectroscopy\cite{kl}.  The nature of correlations in other approaches to exotic hadron spectroscopy, such as the large $N_{c}$ formalism of Jenkins and Manohar\cite{jmexotics} or the chiral soliton model\cite{man}, is less obvious.  Perhaps the spectroscopy considered here could be formulated from those perspectives as well.

Consider the spectrum of hadrons created by light ($u$, $d$, $s$) quark and antiquark sources neutralized with respect to color by a non-dynamical Wilson line in the $(3_{c},\overline 3_{c})$ representation,
\begin{equation}
\label{sources}
C(T) \propto \<0|{\cal S}_{i}^{\dagger}(0,T){\cal P}\left[\exp{i\int_{(0,0)}^{(0,T)} d\xi_{\mu} {\cal 
A}^{\mu}}\right]^{i}_{j}{\cal S}^{j}(0,0)|0\>
\end{equation}
where ${\cal S}^{j}$ is a color triplet product of $u$, $d$, and $s$, quark and/or antiquark fields.  The Wilson line can be regarded as an infinitely heavy, non-dynamical quark, fixed at the origin $x=0$\footnote{More sophisticated lattice-optimized sources can be considered.  One example are the ``all-to-all'' propagators of Ref.~\cite{Foley:2005ac}.  Neutralizing a collection of quarks with a pointlike $\overline 3_{c}$ source is like neutralizing electrons by placing them in the field of an infinitely heavy oppositely charged nucleus.  Alternatively, one might try to neutralize them with a uniform color background field, analogous to the uniform positive charge background introduced in simple models of metals.  Unfortunately I do not know how to implement this idea in a non-Abelian gauge theory\cite{jwn}.}   By studying the large-$T$ behavior of the Euclidean-time correlator, $C(T)$, and looking for a plateau in $\ln C$, discrete states can be identified, if they exist.  The color triplet (or antitriplet) sources, ${\cal S}^{i}$, include the obvious examples $q^{i}$ and $\epsilon_{ijk}q^{j}q^{k}$, corresponding to the simplest heavy mesons and baryons, but also less obvious examples like $[qq\bar q]^{i}$, $[qqqq]^{i}$, $[qqq\bar q]_{i}$ and so on.  All these correlators can be studied using lattice QCD methods.  The spectrum of color sextet combinations of quarks could also be studied using these methods, although I do not consider it here.  The simple case  where ${\cal S}^{i}=q^{i}$ corresponds to $q\bar b$ mesons in the limit of infinite $b$-quark mass.  Hadrons with the quantum numbers created by ${\cal S}^{i}$ can readily be studied in quark models as well.  

Usually we demand that the system we study be accessible to experiment.   However QCD is certainly the right theory of hadrons, and lattice QCD methods will eventually, if not already, provide data that can stand in stead0 of experiment in cases where experiment is impractical.  Non-singlet spectroscopy is such a case.  Although the 
charm quark is heavy, it is not heavy enough to ignore its ${\cal O}(\Lambda/m_{c})$ color and spin dependent interactions with light quarks.  The bottom quark is closer to the ideal, but the spectrum of bottom mesons and (especially) baryons is not very well known.  Furthermore, experiments are expensive and the present funding climate seems to favor ambitious lattice simulations over new experimental initiatives.  In short I proposal to dispense with the usual comparison with experiment, and instead study features of QCD by comparing phenomenological models directly with lattice calculations.  Of course the day may come when the spectrum of $B$-baryons and mesons becomes known in detail.  Then the results of lattice and model calculations can be compared with yet a third approach to QCD:  experiment.

The program described here has its roots in the study of charm exotic mesons and baryons.  The first proposal that $qq \bar q \bar c$ states might be anomalously stable was made by Lipkin in 1977 \cite{Lipkin:1977ie}.  To the extent that the dynamics of the heavy $c$-quark can be ignored, Lipkin's exotic mesons coincide with the $qq\bar q$ states studied here.  Exotic charm baryons were also discussed\cite{baryons}.
  
Early in the study of multiquark hadrons, Mulders, Aerts, and de Swart cataloged the quantum numbers of color ${3}_{c}$ and $\overline {3}_{c}$ combinations of light quarks and antiquarks \cite{Mulders:1978cp}\footnote{I am indebted to D.~Boer for pointing out this reference.  Another, even earlier, but less complete compendium can be found in the appendix to Ref.~\cite{Chan:1978nk}.
See also Refs.~  \cite{Hogaasen:1978jw,deCrombrugghe:1978hi} for studies of $q^{4}\bar q$ from which $q^{2}\bar q$ properties can be inferred.}. They also evaluated the matrix elements of an effective colormagnetic (one gluon exchange) Hamiltonian.  Many of the results described here can be found among the tables of Ref.~\cite{Mulders:1978cp} (see especially Tables II, III, and IV).

The most convenient way to enumerate the quantum numbers of the lightest states composed of $u$, $d$, and $s$ quarks is to use the language of the constituent quark model. The lightest baryons ($qqq$) and mesons ($\bar q q$) observed in Nature correspond to the states that can be built from a collection of quarks and antiquarks, all in the same $j=1/2$ orbital of some mean field.  States fall into irreducible representations of $SU(3)_{c}\times SU(3)_{f}\times  SU(2)_{s}$ (in the $m_{s}=0$ limit).  This elementary approach captures the pseudoscalar and vector nonets of light mesons and the familiar positive parity nucleon octet and the spin-$3/2$ decuplet of baryons.  These are by far the most prominant light hadrons, and there is every reason to expect the approach to be as useful for color non-singlet sectors.  The broad features of the spectrum that results are likely to be more general than the model.  Also the enumeration of states can be translated into an enumeration of local sources for use in lattice calculations.

The object of a lattice program would be to look for exceptionally stable color triplet configurations and to try to understand the stability in terms of quark correlations.  In contrast to the situation among $[q]^{{3}_{c}}$ and $[qq]^{{3}_{c}}$ states, the great majority of the $[qq\bar q]^{{3}_{c}}$, $[qqqq]^{{3}_{c}}$, {\it etc.\/}, states should be unstable against decay into a quark and a color singlet hadron.  Thus $[qq\bar q]^{3_{c}}\to [q\bar q][q] ^{3_{c}}$, $[qqqq]^{{3}_{c}}\to [qqq][q]^{3_{c}}$, and so forth.  There is no quark pair-creation barrier to inhibit these decays.  So if they are kinematically allowed, the lattice correlator would not show a discrete state at all, but merely the $[q\bar q]q$ or $[qqq]q$ continuum.  Thus it would be wise to look first at the channels where strong attractive correlations are indicated by simple models (see below), namely the flavor triplet spin-0 (and spin-1) $[qqqq]^{{3}_{c}}$ and the flavor triplet and antisextet $[qq\bar q]^{{3}_{c}}$ channels.  Lattice studies of these states may depend sensitively on quark mass.  Color triplet $[qq\bar q]$ and $[qqqq]$ states that are in the continuum for the physical quark masses, may move below threshold if quark masses are so large that the pion is not near its physical mass.  This phenomenon is familiar from lattice studies of the $\rho$-meson, which is stable against decay into two pions in many lattice studies.  Some years ago Alford and I used the quark mass to stabilize $[qq\bar q\bar q]$ states in an effort to study mesons composed of more than $q \bar q$\cite{Alford:2000mm}.  The same method may provide an additional degree of freedom in the study of non-singlet spectroscopy.  Remember that having abandoned comparison with experiment, it is no longer necessary to restrict lattice studies to ``physical'' values of quark masses.  Insight can be gained by comparing lattice studies with unphysical quark masses to the predictions of phenomenological models in the same modified world.

In Section 2 I define as simple and compact a notation as possible for states made of many quarks and antiquarks.  I also review two elementary models used to get a first estimate of the spectrum.  In Section 3 I construct the color triplet quark states for $[qq]$, a pedagogical example, $[qqqq]$, and finally $[qq\bar q]$, and discuss the spectrum in the color magnetic and schematic diquark models.  I want to stress that these are strawmen --- merely examples of the kind of phenomenological analyses that could be carried out.  I have omitted some ingredients, like $SU(3)$-flavor symmetry breaking, that are no doubt important, because they require more detailed calculations.  The paper closes with some discussion in 
Section 4.

	The most prominent $[qqqq]^{{3}_{c}}$ states are a flavor triplet multiplet, with the flavor structure, $[[u,d],[d,s]]$, $[[d,s],[s,u]]$, and $[[s,u],[u,d]]$, where the $[\ ,\ ]$ notation denotes flavor antisymmetrization\cite{Stewart:2004pd}.  This multiplet is selected by both the schematic diquark model and the colormagnetic Hamiltonian (though the latter predicts both $J=0$ and $J=1$, while the former prefers $J=0$ only).  Although the quantum numbers are the same, the internal substructure of these states differ significantly in the two models.  In particular, the colormagnetic ground state has a much richer diquark structure including color and flavor sextet diquarks.  In contrast, by hypothesis, the ground state of the schematic diquark model includes color and flavor antitriplet diquarks only.  The same situation  arises in the $[qq\bar q]^{{3}_{c}}$ sector.  Both models favor a spin-1/2 (odd parity) flavor triplet plus antisextet ground state.  But the models differ sharply on the diquark content of the state.  If $[qqqq]^{{3}_{c}}$ or $[qq\bar q]^{{3}_{c}}$ states can be found in lattice computations, the next step would be to probe their diquark content using suitable local currents.

\section{Preparation}

\subsection{Notation and basis states}

The light, $u$, $d$ and $s$, quarks carry color ($SU(3)_{c}$), flavor ($SU(3)_{f}$), and spin ($SU(2)_{s}$) labels.  In this paper except for an occasional comment, I ignore the  $SU(3)_{f}$ symmetry breaking driven by the strange quark mass.  Here I consider only states composed of quarks and antiquarks all in the same mode of some mean field (generated by the pointlike, static, infinitely massive $\overline 3_{c}$ source) assumed to have total angular momentum-1/2, since these are likely to be the lightest states\footnote{Although I use the term ``spin'', my analysis \emph{is not} restricted to non-relativistic quarks because the rotations of relativistic quarks in the state of total angular momentum $1/2$ are also described by an $SU(2)$ algebra.}.  Therefore the quarks (and antiquarks) must be treated as \emph{identical particles} save their color, flavor, and spin indices.  In the language of atomic or nuclear spectroscopy  such quarks are  ``equivalent'' particles.

In addition it will be very useful to introduce \emph{colorspin}, the $SU(6)_{cs}$ generated by the the combination of color and spin transformations.  This should not be confused with the old quark model $SU(6)_{f\!s}$ composed of {\it flavor\/}$\ \times$ spin.  All irreducible representations (irreps) will be labeled by their dimension.  Even though this is occasionally ambiguous, it is more familiar ({\it viz.\/} the ``octet'' versus ``the $\{2,1\}$'').  Ambiguities are resolved by reference to Young diagrams.  Since both color and (light quark) flavor are $SU(3)$ symmetry groups, I use a subscript $c$ or $f$ to distinguish between them.  For further clarity, I use boldface for flavor-$SU(3)_{f}$. $SU(6)_{cs}$ representations will be  in brackets.  Thus the state of a single quark is labeled
\begin{equation}
|q,[6] (3_{c},2)\ \mathbf{3}_{f}\>
\label{quark}
\end{equation}
Note that the color and spin labels are grouped together following the colorspin because the $SU(3)_{c}\times SU(2)_{s}$ subgroup content of a $SU(6)_{cs}$ irrep will often be important.  Occasionally I refer to irreps by their corresponding Young diagrams.

The states of two quarks --- diquarks --- need special attention and notation.  Two equivalent quarks must be antisymmetric under combined colorspin and flavor transformations.  They can be symmetric (~$\tiny\yng(2)$~) in colorspin, $[21]$, and antisymmetric (~$\tiny\yng(1,1)$~) in flavor, $\mathbf{\overline 3}_{f}$, or antisymmetric in colorspin, $[15]$, and symmetric in flavor, $\mathbf{6}_{f}$.  The $[21]$ contains $(6_{c},3)$ and $( {\overline 3}_{c},1)$.  The $[15]$
contains $(6_{c},1)$ and $({\overline 3}_{c},3)$.  So a complete specification of two quarks (in the lowest $j=1/2$ mode) can be written\footnote{I have suppressed the ``magnetic'' quantum numbers like $m_{s}$, $I$, $I_{3}$, $Y$, and the analogous color labels.},
\begin{eqnarray}
\label{quarkquark}
|\alpha\>&\equiv& |qq,[21](\overline 3_{c},1)
\ \mathbf{\overline{3}}_{f}\>\nonumber\\
|\beta\>&\equiv&|qq,[15](\overline{3}_{c},3)
 ,\mathbf{6}_{f}\>\nonumber\\
|\gamma\>&\equiv&|qq,[15]( {6}_{c},1)
 ,\mathbf{6}_{f}\>\nonumber\\
|\delta\>&\equiv&|qq,[21](6_{c},3)
\ \mathbf{\overline{3}}_{f}\>
\end{eqnarray}
The reader will recognize the ``good'' and ``bad'' diquarks of Ref.~\cite{jw}, both of which are color anti-triplets, as $|\alpha\>$ and $|\beta\>$ respectively.  The color sextet diquarks were not considered in most recent discussions of exotic spectroscopy, although they do play an important role in the triquark picture of Karliner and Lipkin\cite{kl}.  Anti-diquark states are constructed analogously.  

It is easy to write down local composite operators that can create these states from the vacuum.  For example, the $\alpha$ diquark couples to
\begin{equation}
\label{operators}
	[qq]_{ia}=\epsilon_{ijk}\epsilon_{abc
	}(i\sigma_{2})_{\alpha\beta}\ 
	q^{jb}_{\alpha}q^{kc}_{\beta}=\epsilon_{ijk}\epsilon_{abc}\ \overline
	q^{jb}_{\rm C}\gamma_{5}q^{kc}
\end{equation}
where $ijk$ ($abc$) are color (flavor) labels and $\overline q_{\rm C}=-iq^{T}\sigma^{2}\gamma_{5}$.  Note the $\gamma_{5}$ in eq.~(\ref{operators}) ensures that the diquark operator has \emph{positive} parity.  Another example: the color-sextet, spin-one ($\delta$) diquark couples to the operator
\begin{equation}
\label{operators2}
	[qq]^{\{ij\},r}_{a}=s^{ij}_{k\ell}\epsilon_{abc} \ \overline
	q^{kb}_{\rm C}\gamma^{r}\gamma_{5}q^{\ell c}
\end{equation}
where $s^{ij}_{k\ell}=\delta^{i}_{k}\delta^{j}_{\ell}
+\delta^{i}_{\ell}\delta^{j}_{k}-\frac{2}{3}\delta^{ij}\delta_{k\ell}$ projects on the $6_{c}$ irrep, and $\gamma^{r}$ selects spin-one.  In general there will be more than one local composite operator corresponding to each color non-singlet $q^{n}\overline q^{m}$ state.  Note that odd parity operators like $\epsilon_{ijk} \epsilon_{abc}\ \overline q^{jb}_{\rm C}q^{kc}$ create quark model states in which one quark must be excited.  They are presumed to be much heavier and are not considered further here.

When quarks are combined with antiquarks there are two complete sets of commuting operators to describe their color $\times$ spin $\times$ flavor wavefunction that are useful.  The choice hinges on the fact that the quadratic Casimir operator for colorspin does not commute with the quadratic Casimir operators for the spin and color of the quarks and antiquarks separately.  One choice is to forget total colorspin and label irreps by the color and spin of the quarks and antiquarks.  An example from the world of $q^{2}\bar q^{2}$-states would be $|\alpha\overline\alpha\>$, a color singlet, spin-0 meson made exclusively of a good diquark and good antidiquark   \cite{Maiani:2004uc}.  This is not a state of definite total colorspin.  The second choice is to diagonalize total colorspin, in which case the quark and antiquark spin and color are no longer definite\cite{rjmulti}.  For example, a $|q^{2}\bar q^{2}\>$ colorspin singlet is a linear superposition of $|\alpha\bar\alpha\>$, $|\beta\bar\beta\>$, {\it etc.\/}

\subsection{Interactions}

As emphasized in the Introduction I am not advocating any particular model for interquark interactions.  Instead I want to illustrate the capacity for color non-singlet spectroscopy to probe quark correlations and distinguish between models.  

The schematic diquark model is so simple that it requires little explanation:  Among the four diquarks, the color sextets $\gamma$ and $\delta$, are ignored --- assumed to be heavy enough that states in which they appear are grossly unstable and could not be distinguished from the continuum.  This leaves the color antitriplets, $\alpha$ and $\beta$.  $\alpha$, the ``good'' diquark is assumed to be about 200 MeV lighter than $\beta$, the ``bad'' diquark.  The $\alpha$ and $\beta$ diquarks are assumed to be sufficiently strongly correlated that the spin interactions between quarks in different diquarks can be ignored.  Nothing is said about the spin interactions between two $\beta$-diquarks, {\it etc.\/}  This is a crude model, good perhaps as a qualitative guide.  It does remarkably well in the  $qqq$, and $\bar q \bar q qq$ sectors, where it reproduces the ordering of the lightest states.  A more sophisticated schematic diquark model can reproduce much of the spectrum of excited baryons\cite{Wilczek:2004im,selem}

The colormagnetic interaction model, on the other hand, is more precisely defined.
The transverse gluon exchange force between equivalent $j=1/2$ quarks and antiquarks was first considered in Ref.~\cite{DeRujula:1975ge}, where it was applied to the spectrum of $q\bar q$ and $qqq$ eigenstates.  Later in Refs.~\cite{rjmulti}, it was applied to multiquark states.  The Hamiltonian is
\begin{equation}
\label{hcs}
{\cal H}_{cs} = -\frac{1}{4}{\cal M} \sum_{i>j} \vec \sigma_{i} \cdot
\vec\sigma_{j}  \ \ \tilde\beta_{i}  \cdot\tilde\beta_{j} 
\end{equation}
Here $\{\half\sigma_{j}^{k}; k=1,2,3\}$ are the three generators of the spin rotations of the $j^{\rm th}$ quark represented by Pauli matrices, and normed by ${\rm Tr}(\sigma^{k})^{2}=2$, and  $\{\half\beta_{j}^{\ell}; \ell=1,2,... 8\}$ are the eight generators of the color transformations of the $j^{\rm th}$ quark, represented by the Gell-Mann matrices, and normed the same way.  The sum ranges over all quarks and antiquarks\footnote{For antiquarks $\overline\sigma^{k} = -[\sigma^k]^{*}$ and 
$\overline\lambda^{\ell}=-[\lambda^\ell]^{*}$.}.  ${\cal M}$ is a model-dependent reduced matrix element parameterizing the strength of the interaction.  ${\cal M}$ presumably depends on the quark content of the state, so we label it ${\cal M}_{qq}$, ${\cal M}_{qqqq}$, or ${\cal M}_{qq\bar q}$ as appropriate.  ${\cal M}$ also depends on light quark masses, so if $SU(3)_{f}$ violation were included, ${\cal M}$ would have to appear inside the summation.  It is important to remember that this analysis is not limited to non-relativistic quarks and antiquarks.  Eq.~(\ref{hcs}) is the spin-dependent part of the relativistic, gauge invariant interaction $\overline q\gamma_{\mu} {\beta^{k}} q\   {D}^{\mu\nu}_{k\ell}\ 
\overline q\gamma_{\nu}{\beta^{\ell}}q$ projected onto the $j=1/2$ sector\cite{DeGrand:1975cf}. 

 The matrix elements of ${\cal H}_{cs}$ can be re-expressed in terms of the quadratic Casimir operators of the color, spin, and colorspin of the quarks and antiquarks\cite{rjmulti},
\begin{eqnarray}
\label{cscasimir}
{\cal H}_{cs} &=& {\cal M} \left(2 N +\half {\cal C}_{cs}(\rm TOT)-\frac{1}{3}S_{\rm TOT}(S_{\rm TOT}+1)-\half {\cal C}_{c}({\rm TOT})
\right.\nonumber\\
&+&\left. {\cal C}_{c}(Q)+\frac{2}{3}S_{Q}(S_{Q}+1)-{\cal C}_{cs}(Q)
+{\cal C}_{c}(\overline Q)+\frac{2}{3}S_{\overline Q}(S_{\overline Q}+1)-{\cal C}_{cs}(\overline Q)
\right)
\end{eqnarray}
${\cal C}_{c}$ and ${\cal C}_{cs}$ are the quadratic Casimir operator eigenvalues for the color and colorspin irreps of the quarks ($Q$), antiquarks ($\overline Q$), and total system (TOT) respectively. $N$ is the total number of quarks and antiquarks in the state.  The Casimir operators are normalized conventionally \cite{conventional}, so, for example ${\cal C}_{c}( {3})=4/3$.  The Casimir operator eigenvalues for the relevant irreps of $SU(3)$ and $SU(6)$, together with other important information, are listed in Tables \ref{table1} and \ref{table2}.  In most cases the colorspin interaction is diagonal in the states with definite colorspin, color, and spin for quarks, antiquarks, and the total system.  In two cases the interaction mixes states of different total colorspin.  Examples will clarify the situation. 
\begin{table}
\caption{$SU(3)$ representations relevant for both color and flavor. In the weight diagrams isospin multiplets (with the same baryon number, spin, and parity) shown in the same color can mix when $SU(3)_{f}$ symmetry is broken. Multiplets in black do not mix.}
\begin{tabular}{|c|c|c|}
\hline
$SU(3)_{c}$ Irrep &
      ${\cal C}_{c}$& Weight Diagram\\
\hline
\parbox{1in}{\begin{center}$ {\mathbf{3}}_{f}\leftrightarrow \tiny\yng(1)$ \end{center}} & $\frac{4}{3}$& \includegraphics[width=1cm]{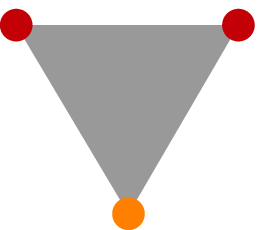}
\\ 
\hline
\parbox{1in}{\begin{center}\vskip -40pt ${{\overline{\mathbf{6}}}_{f}\leftrightarrow \tiny\yng(2,2)} $\end{center}}  &\parbox{.25in}{\begin{center}\vskip -40pt $\frac{10}{3}$\end{center}}& 
\includegraphics[width=2cm]{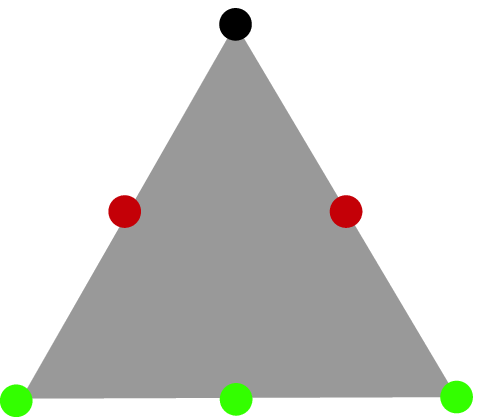}\\
\hline
\parbox{1in}{\begin{center}\vskip -55pt
${\mathbf{15}}_{f}\leftrightarrow \tiny\yng(3,1)$ \end{center}}&\parbox{.25in}{\begin{center}\vskip -55pt $\frac{16}{3}$
\end{center}}& \includegraphics[width=2.5cm]{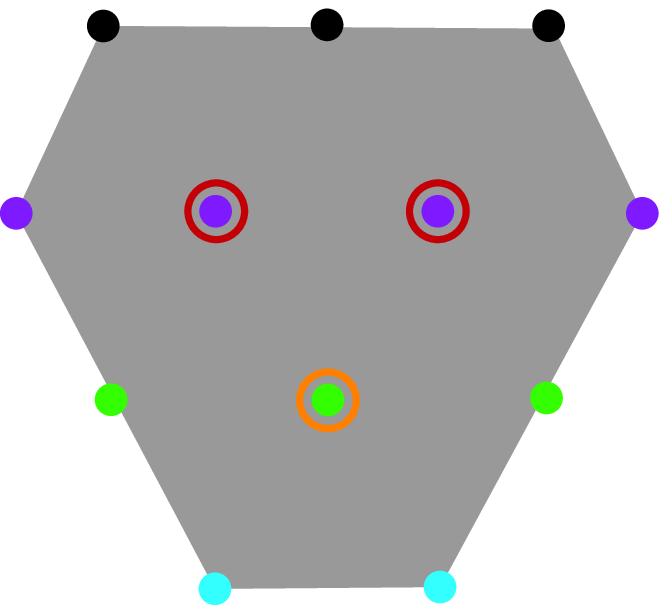}\\
\hline
\parbox{1in}{\begin{center}\vskip -60pt
${\mathbf{15'}}_{f}\leftrightarrow\tiny\yng(4) $\end{center} }&
 \parbox{.25in}{\begin{center}\vskip -60pt $\frac{28}{3}$
 \end{center}} & \includegraphics[width=3cm]{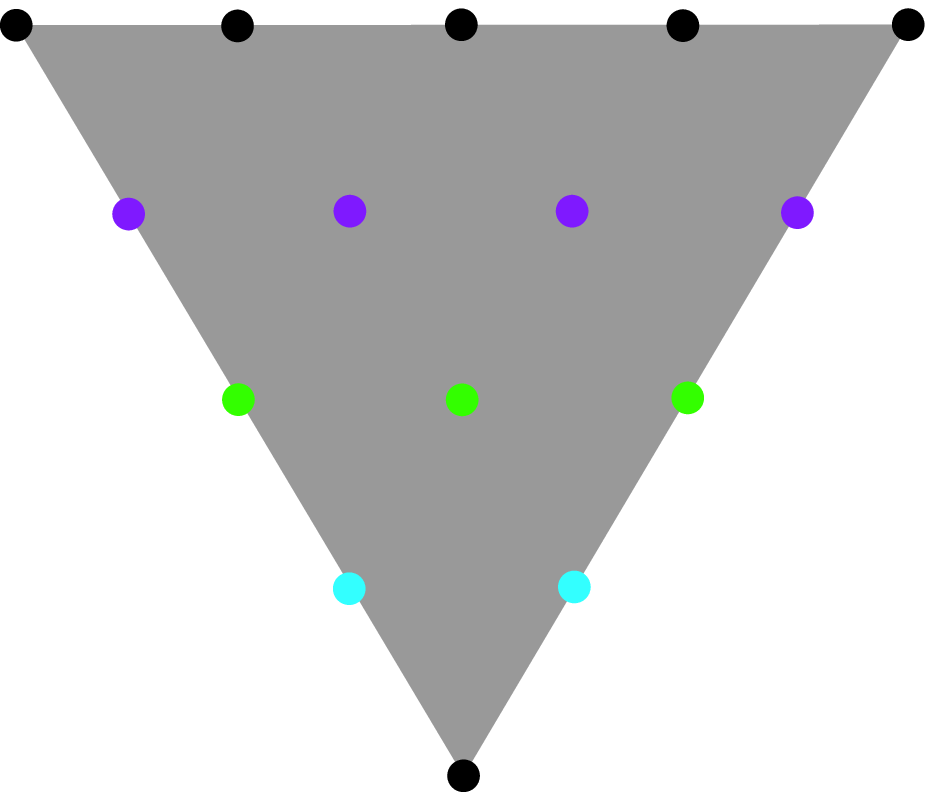}\\
\hline
\end{tabular}
\label{table1}
\end{table}

\begin{table}
\caption{$SU(6)_{cs}$ Representations}
\begin{tabular}{|c|c|c|c|}
\hline
 $SU(6)_{cs}$ Irrep &
      ${\cal C}_{cs}$ & $SU(3)_{c}\times SU(2)$  Decomposition  & $SU(3)_{f}$ Irrep\\
\hline
\multicolumn{4}{|c|}{ Representations relevant to $q$}\\
\hline
$[6]\leftrightarrow\tiny\yng(1)$     & $\frac{35}{6}$& $(3_{c},2)$& ${\mathbf 3}_{f}\leftrightarrow$\ \tiny\yng(1)\\ 
\hline
\multicolumn{4}{|c|}{ Representations relevant to $q^{2}$} \\
\hline
$[21]\leftrightarrow\tiny\yng(2)$ & $\frac{40}{3}$&
 $ (\overline 3_{c},1) ,(6_{c},3)$& 
 $\overline{\mathbf 3}_{f}\leftrightarrow \ \tiny\yng(1,1)$
 \\
\hline
$[15]\leftrightarrow\tiny\yng(1,1)$
  & $\frac{28}{3}$&
 $(\overline 3_{c},3), (6_{c},1) $& 
  $ {\mathbf 6}_{f}\leftrightarrow \ \tiny\yng(2)$\\
\hline
\multicolumn{4}{|c|}{ Representations relevant to $q^2 
\overline q $ }\\
\hline
$[120]\leftrightarrow\tiny\yng(3,1,1,1,1)$  & $\frac{119}{6}$&
\parbox{1.8in}{\begin{center}$(3_{c},2),(3_{c},4),
(\overline 6_{c},2)$\\ $(15_{c},2),(15_{c},4)$\end{center}}& 
 $ {\mathbf 3}_{f}\oplus \overline{\mathbf{6}}_{f}\leftrightarrow \ \tiny\yng(1)\oplus \tiny\yng(2,2)$
\\
\hline
$[84]\leftrightarrow\tiny\yng(2,2,1,1,1)$  & $\frac{95}{6}$&
\parbox{1.8in}{\begin{center}($\overline 3_{c},2), (\overline 3_{c},4)$\\ $
(6_{c},2), (6_{c},4), (15_{c},2)$\end{center}}&
$ {\mathbf 3}_{f}\oplus {\mathbf{15}}_{f}\leftrightarrow \ \tiny\yng(1)\oplus \tiny\yng(3,1)$
\\
\hline
\multicolumn{4}{|c|}{ Representations relevant to $q^4 $}\\
\hline
$[\overline{15}]\leftrightarrow\tiny\yng(1,1,1,1)$ & $\frac{28}{3}$&
\parbox{1.8in}{\begin{center}$(3_{c},3),(\overline 6_{c},1) $\end{center}}& $ {\mathbf{15}}'_{f}\leftrightarrow \ \tiny\yng(4)$ \\
\hline
$[105]\leftrightarrow\tiny\yng(2,1,1)$  & $\frac{52}{3}$&
\parbox{1.8in}{\begin{center}$(3_{c},1),(3_{c},3),( 3_{c},5)$\\ $(\overline 6_{c},3),(15_{c},1),(15_{c},3)$\end{center}}& 
$ {\mathbf{15}}_{f}\leftrightarrow \ \tiny\yng(3,1)$
\\
\hline
$[105']\leftrightarrow\tiny\yng(2,2)$ 
 & $\frac{64}{3}$&
\parbox{1.8in}{\begin{center}
$ (3_{c},3),(\overline 6_{c},1),(\overline 6_{c},5)$ \\ $(15_{c},3),(15'_{c},1)$\end{center}}& 
$ {\overline{\mathbf 6}}_{f}\leftrightarrow \ \tiny\yng(2,2)$
\\
\hline
$[210]\leftrightarrow\tiny\yng(3,1)$ 
 & $\frac{76}{3}$&
\parbox{1.8in}{\begin{center}$(3_{c},1), (3_{c},3),(\overline 6_{c},3)$\\ $(15_{c},1), (15_{c},3), (15_{c},5), (15'_{c},3)$\end{center}}& 
$ {\mathbf 3}_{f}\leftrightarrow \ \tiny\yng(1)$
\\
\hline
\end{tabular}
\label{table2}
\end{table}%
 
\section{Classification of Color Non-Singlet States}

The goal is to construct the spectra of the constituent quark model in the $[q^{2}\overline q]^{3_{c}}$ and $[qqqq]^{3_{c}}$ sectors.  First, for orientation, I illustrate the methods for $[qq]^{\overline 3_{c}}$.  Next I consider $[qqqq]^{3_{c}}$, and finally $[q^{2}\overline{q}]^{3_{c}}$, which is slightly more complicated.  I follow the same program in all three cases:  First  enumerate the states and their quantum numbers.  Next illustrate model predictions by constructing ``strawman'' spectra in the colormagnetic and diquark models.  Finally   discuss the spectra and what theoretical issues they illustrate. 

\subsection{I.\quad $[qq]^{\overline 3_{c}}$}

Color antitriplet $qq$-states correspond  to   heavy quark baryons in the approximation that the residual, {\it e.g.\/} spin-dependent, forces between the light quarks and the heavy quark can be ignored.  This is the $M\to\infty$ limit of heavy quark effective field theory\cite{MW}.  Of course the results are well known and are included here only to set the stage for more complicated calculations.  

First, construct properly antisymmetrized states of two equivalent quarks by combining colorspin $\times$ flavor irreps.  This was done in the previous section.  The results are summarized in eq.~(\ref{quarkquark}) and in lines-2 and 3 of Table~\ref{table2}.
The spectrum of color antitriplet states consists of the $\alpha$ and $\beta$-diquarks which are $\overline{\mathbf 3}_{f}$ and $\mathbf 6_{f}$ respectively.  The $SU(3)_{f}$ weight diagrams for these multiplets are shown in the upper half of  Fig.~\ref{multiplets}.  None of these states mix even in the presence of $SU(3)_{f}$ violation because the multiplets with the same isospin and hypercharge have different total angular momentum.

In the colormagnetic model the  colorspin interaction matrix elements (which we call $\Delta$ following Ref.~\cite{Mulders:1978cp}) are determined by eq.~(\ref{cscasimir}), 
\begin{equation}
\begin{array}{lll}
\Delta(\alpha)\equiv\Delta([21](\overline{3}_{c},1)) &=& -2{\cal M}_{qq}\nonumber\\
\Delta(\beta)\equiv\Delta([15](\overline{3}_{c},3)) &=& +\frac{2}{3}{\cal M}_{qq}.
\end{array}
\end{equation}
As expected, the $\beta$ diquark is heavier, as assumed in the schematic diquark model.  
%
%
  
The phenomenology of these states is well known from the analogs in the charm sector.  They combine with the charm quark to make the lightest positive parity charm baryons.  The flavor  hypercharge ($Y$) $2/3$ state in the $\overline \mathbf{3}_{f} $ joins the charm quark to make the spin-$1/2$ $\Lambda_{c}$.  The $Y=-1/3$ isodoublet in the $\overline \mathbf{3}_{f} $ couples to $c$-quark to make spin-$1/2$ $\Xi_{c}$ states.  However the spin-triplet  $Y=-1/3$ isodoublet in the flavor $\mathbf{6}_{f}$ can also join the $c$-quark to make a spin-$1/2$ $\Xi_{c}$ state.  When the strange quark mass is ``turned on'' these
two states mix to form the $\Xi_{c}$ and the $\Xi'_{c}$.  Note that the $\alpha$ and $\beta$ diquarks \emph{do} mix within the $I=1/2$ $qqc$ baryons because the total spin of the states is the same.  This mixing vanishes in the $m_{c}\to \infty$ limit which defines color non-singlet spectroscopy.  The $Y=2/3$ isovector in the $\mathbf{6}_{f}$ couples to the $c$-quark to make the spin-$1/2$ $\Sigma_{c}$ and the spin-$3/2$ $\Sigma^{*}_{c}$.  The remaining charm baryons are the spin-$1/2$ and $3/2$ $\Omega_{c}$ and the spin-$3/2$ $\Xi_{c}$ composed of flavor $\mathbf{6}_{f}$ diquarks coupled to the charmed quark.  Taking linear combinations of masses to eliminate the spin interactions with the charm quark, it is possible to isolate the $\alpha-\beta$ diquark mass difference for the non-strange quarks \cite{Jaffe:2004ph}, 
\begin{equation}
	\Delta(\beta)-\Delta(\alpha) = \frac{8}{3}{\cal M}_{qq} \approx 212\ \hbox{MeV}
\label{goodbad}
\end{equation}
so for $u$ and $d$ quarks, ${\cal M}_{qq}\approx 80$ MeV.

Note that the color sextet diquarks play no role in $[qq]^{\overline 3_{c}}$ spectroscopy.  Though one should take care to distinguish the simple classification language of quark models from the gauge invariant description appropriate to lattice QCD.  Thus a source with the quantum numbers of a color antitriplet diquark can mix with operators in which the quarks are in a color sextet and the overall color is restored to $\overline 3_{c}$ by a gluon.  

\subsection{II.\quad $[qqqq]^{3_{c}}$}

The possible states of four equivalent quarks are highly constrained by Fermi statistics.  The four possible colorspin multiplets are ennumerated in the last four rows of Table~\ref{table2}\footnote{The totally symmetric colorspin configuration ($\tiny\yng(4)$) is excluded because four quarks cannot be antisymmetrized over three flavors}.  They are associated with the flavor irreps $\mathbf{15'}_{f}$, $\mathbf{15}_{f}$, $\mathbf{\overline{6}}_{f}$, and $\mathbf{3}_{f}$ respectively. 
\begin{floatingfigure}{.65\textwidth} 
\includegraphics[width=9cm]{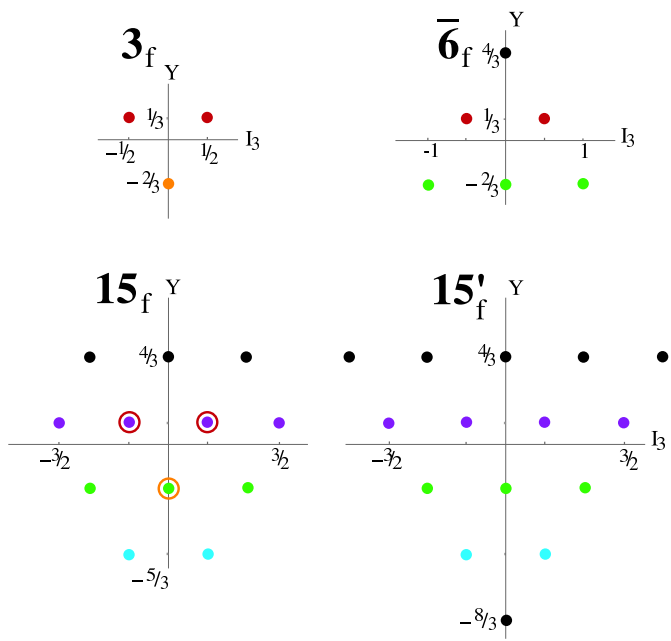}
\caption{\small $SU(3)_{f}$ diagrams for the $[qq]^{3_{c}}$, $[qqqq]^{{3}_{c}}$ and $[qq\bar q]^{{3}_{c}}$ states.  States with the same isospin and hypercharge (labeled in a common color) mix when $SU(3)_{f}$ violation is turned on if they have the same total angular momentum} \label{multiplets}
\end{floatingfigure}  
\noindent
 Only color triplet states are of interest.  The spin and flavor irreps are summarized in Table~\ref{table3} along with other spectroscopically important information.  The $SU(3)_{f}$ multiplets are shown in Fig.~\ref{multiplets}.  States that mix when $SU(3)_{f}$ symmetry is broken can be read off from the table and the figure.  For example from Table~\ref{table3} we see that there are two $J=0$ multiplets one  $\mathbf{3}_{f}$ and one $\mathbf{15}_{f}$.  From Fig.~\ref{multiplets} we see that the isodoublet  with $Y=1/3$ appears in both multiplets, as does the isosinglet with $Y=-2/3$.  These pairs of states will mix when $SU(3)_{f}$ violation is turned on.  All these states are ``exotic'' --- they cannot mix with and therefore be confused with any states with a smaller number of quarks.  

\begin{table}
\caption{$[qqqq]^{3_{c}}$ States in the Colorspin Basis}
\begin{tabular}{|c|c|c|c|cccc|}
\hline
 $SU(6)_{cs}$ Irrep &
      Spin & $SU(3)_{f}$ Irrep & $\Delta$&$\alpha$&$\beta$&$\gamma$&$\delta$\\
\hline
$[\overline{15}]$& $1$ & ${\mathbf{15'}_{f}}$ & $\frac{14}{3}{\cal M}_{qqqq}$&
$0$&$\frac{2}{3}$&$\frac{1}{3}$&$0$\\
\hline

$[105]$& $2$ & ${\mathbf{15}_{f}}$ & $2{\cal M}_{qqqq}$&
$0$&$\frac{2}{3}$&$0$&$\frac{1}{3}$\\
\hline
$[105]$& $1$ & ${\mathbf{15}_{f}}$ & $\frac{2}{3}{\cal M}_{qqqq}$&
$\frac{1}{6}$&$\frac{1}{2}$&$\frac{1}{6}$&$\frac{1}{6}$\\
\hline
$[105]$& $0$ & ${\mathbf{15}_{f}}$ & $0$&
$\frac{1}{4}$&$\frac{5}{12}$&$\frac{1}{4}$&$\frac{1}{12}$\\
\hline
$[105']$& $1$ & ${\mathbf{\overline{6}}_{f}}$ & $-\frac{4}{3}{\cal M}_{qqqq}$&
$\frac{1}{4}$&$\frac{5}{12}$&$\frac{1}{12}$&$\frac{1}{4}$\\
\hline
$[210]$& $1$ & ${\mathbf{3}_{f}}$ & $-10/3{\cal M}_{qqqq}$&$\frac{1}{3}$&$\frac{1}{3}$&$0$&$\frac{1}{3}$\\
\hline
$[210]$& $0$ & ${\mathbf{3}_{f}}$ & $-4{\cal M}_{qqqq}$&$\frac{5}{12}$&$\frac{1}{4}$&$\frac{1}{12}$&$\frac{1}{4}$\\
\hline
\end{tabular}
\label{table3}
\end{table}%

It is easy to work out the matrix elements of the colormagnetic interaction from eq.~(\ref{cscasimir}), remembering that in this case there are no antiquarks, so the quark and ``total'' Casimirs are to be identified\cite{Mulders:1978cp}.  The results are given in Table~\ref{table3} and shown graphically in Figure~\ref{qqqq}.  In the $SU(3)_{f}$ limit there are seven distinct multiplets.  
\begin{floatingfigure}{.65\textwidth} 
\includegraphics[width=9cm]{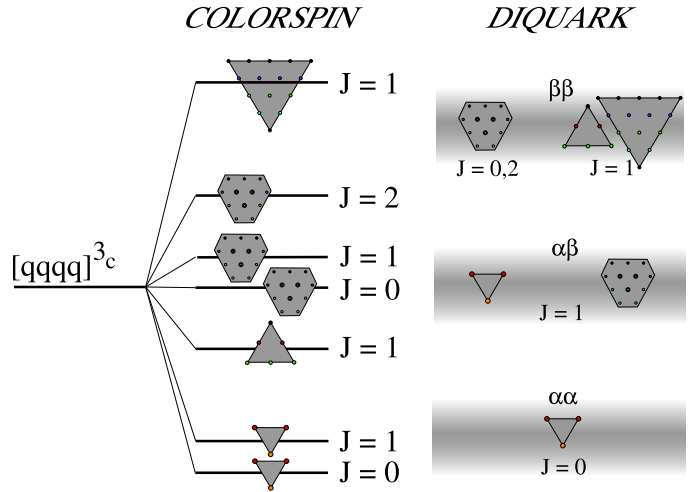}
\caption{\small  The spectrum of $[qqqq]^{{3}_{c}}$ states.  The first column shows colormagnetic model splittings.  The second gives a sketch of the ordering of states in the schematic diquark model.  $SU(3)_{f}$ multiplets are labeled by icons of their weight diagrams.}
  \label{qqqq}
  \vspace*{.2in}
\end{floatingfigure}

The schematic diquark model builds $[qqqq]^{3_{c}}$ states out of pairs of correlated color $\overline 3_{c}$ diquarks, $|\alpha\rangle \equiv |qq,(\overline 3_{c},2)\mathbf{\overline{3}}_{f}\rangle$ and $|\beta\rangle \equiv|qq,(\overline{3}_{c},2)\mathbf{6}_{f}\rangle$. Both $|\alpha\rangle$ and $|\beta\rangle$ are bosons.  States involving color sextet diquarks are assumed to be much heavier and grossly unstable against decay into $[qqq]q^{3_{c}}$.  The allowed $[qq][qq]$ states are enumerated in Table~\ref{table4}.  Certain states, like the $\mathbf{\overline{6}}_{f}$ in $|\alpha\alpha\>$ are forbidden by bose statistics\cite{jw}.   Note that the totality of states in the schematic diquark model are the same as the colormagnetic model.  Also the candidate for the lightest state is the same:  a spin singlet, flavor $\overline{\mathbf 3}_{f}$.  This is just the state recently discussed by Stewart, Wessling, and Wise\cite{Stewart:2004pd} in connection with bottom and charm exotic baryons.  The quark content of these states was given in the Introduction.   

\begin{table}
\caption{$[qqqq]^{3_{c}}$ States in a schematic diquark model}
\begin{tabular}{|c|c|c|c|}
\hline
  Diquark content &
      Spin & $SU(3)_{f}$ Irrep & ``Mass''\\
\hline
$\alpha\alpha$ & $0$ & $\mathbf{3}_{f}$&$M_{qqqq}$\\
\hline
$\alpha\beta$& $1$ & $ \mathbf{3}_{f}\oplus\mathbf{15}_{f} $ & $M_{qqqq}+ \Delta M_{qqqq}$\\
\hline
$\beta\beta$& $0 \oplus 2$ & ${\mathbf{15}_{f}}$ & $M_{qqqq}+2\Delta M_{qqqq}$\\
\hline 
$\beta\beta$& $1$ & $ \mathbf{\overline{6}}_{f}\oplus\mathbf{15'}_{f}$ & $M_{qqqq}+2\Delta M_{qqqq}$\\
\hline
\end{tabular}
\label{table4}
\end{table}%

The spectrum of $[qqqq]^{3_{c}}$ states in the colormagnetic model is sketched in Fig.~\ref{qqqq} and compared with the spectrum obtained the schematic strongly correlated diquark model.  The ordering of states in the colormagnetic model follows the simple ``Hund's Rule'' of QCD\cite{rjmulti}:  Antisymmetric flavor irreps are favored.  Within a multiplet with the same flavor irrep, low spin is favored.  The diquark scheme yields a similar, but not identical pattern.

The ground state multiplet is the same in both approaches, though the colormagnetic scheme allows $J=1$ as well as $J=0$.  In the schematic diquark model a $J=1$ $\mathbf 3_{f}$ is nearby.  If, as suggested in the Introduction, only the lightest multiplet(s) of $[qqqq]^{3_{c}}$ states are  stable against falling apart into the $[qqq]q^{3_{c}}$ continuum, then it will be very difficult, if not impossible to distinguish these two different pictures on the  basis of their predictions for the spectrum.  

Nevertheless the internal structure of these states is very different in these pictures.  In the schematic diquark model pairs of quarks are correlated into diquarks which are dynamically stable.  Their individual spins and colors are sequestered from residual interactions with the quarks trapped in the other diquark.  In contrast, the colormagnetic model treats all quarks on the same footing --- it is a \emph{symmetric} quark model.  Symmetry alone therefore dictates that \emph{color sextet} as well as color antitriplet diquarks are present in the hadron.  

To quantify the difference I calculate the probability that a given quark in each eigenstate of the colormagnetic Hamiltonian finds itself in each of the four possible diquark states, $\alpha$, $\beta$, $\gamma$, or $\delta$.  The same analysis can be applied to the diquarks that occur in the familiar baryons.  For example, in the symmetric quark model description of the $\Delta$-baryon each quark has unit probability to be found in the $\beta$ (color antitriplet, spin-1) diquark, while in the nucleon each quark has probability 1/2 to be in the $\alpha$ or $\beta$ diquark.  Some simple tricks described in the Appendix can be used to extract the analogous probabilities for $[qqqq]^{3_{c}}$ states.  The  resulting probabilities are given in Table~\ref{table3}.  Note that the lightest state, the $\mathbf{3}_{f}$ with spin zero, is not entirely correlated into $\alpha$ diquarks.  Instead it includes significant admixtures of the ``bad'' diquark, $\beta$, and the color sextet diquark, $\delta$.  These admixtures are required by the fact that the quarks are all in the same spatial state.  In the schematic diquark model the two diquarks are considered sufficiently correlated that antisymmetrization between two diquarks is not an important effect.  If one or more of these states can be identified on the lattice, then the matrix elements of local diquark currents can shed light on this substructure. 

\subsection{III.\quad $[qq\bar q]^{3_{c}}$}

This is the most complicated case considered here.  Two new issues arise: first, construction of states of good total colorspin is more complicated; and second, the colormagnetic interaction mixes states of different colorspin.  Although many of the resulting $[qq\bar q]^{3_{c}}$ states have the same flavor and spin as the quarks themselves, $[qq\bar q]^{{3}_{c}}$ and $[q]^{3_{c}}$ do not mix because they have opposite parity.

 These states can be constructed by coupling the quark pairs up to definite colorspin$\times$flavor, and then coupling the antiquark. All states have \emph{odd parity}.  The classification breaks up naturally according to the colorspin (and flavor) of the quark pairs.  There are two possibilities, either the quarks are symmetric ($[21] = {\tiny\yng(2)}\ $) or antisymmetric ($[15]={\tiny\yng(1,1)}\ $) in colorspin.  In both cases the $qq$ state must be coupled to an antiquark ($[\overline 6]=\overline{\tiny\yng(1)}\ $) to a total colorspin and flavor representation.  The possibilities are
\begin{eqnarray}
\label{symmetric}
\left([21]\ \overline{\mathbf{3}}_{f}\right)\otimes \left([\overline 6]\ {\mathbf{\overline{3}}_{f}}\right)&=&\left([6]\oplus [120]\right)\ \left(\mathbf{3}_{f}\oplus{\overline{\mathbf{6}}_{f}}\right)\nonumber\\
\left([15]\  {\mathbf{6}}_{f} \right)\otimes \left([\overline 6]\ {\mathbf{\overline{3}}_{f}}\right)&=&\left([6]\oplus [84]\right)\ \left(\mathbf{3}_{f}\oplus{ {\mathbf{15}}_{f}}\right)
\end{eqnarray}
 Note that in the $SU(3)_{f}$ limit the states in which the diquarks are in the $[21]_{cs}$ (first row) and $[15]_{cs}$ (second row) do not mix because the quarks are in different irreps of $SU(3)_{f}$.  When $SU(3)_{f}$ violation is added, they can mix.
First I consider the quark pair symmetric in colorspin (thus antisymmetric in flavor) and then the  opposite case.

\bigskip 

1.\quad  $([21]\ \overline{\mathbf{3}}_{f})\otimes ([\overline 6]\ {{\overline{\mathbf 3}_{f}}})$
\medskip

These states are in the $\overline{\mathbf{3}}_{f} \otimes\overline{\mathbf{3}}_{f} \ = \ \mathbf{3}_{f}\oplus\overline{\mathbf{6}}_{f}$ of flavor --- the same multiplets as the diquarks (see Fig.~\ref{multiplets}).  I return to the flavor structure after analyzing their color and spin structure.
The overall colorspin irreps are determined by the Clebsch Gordan series for $[21]\times[\overline 6]$, which includes the fundamental and a 120-dimensional irrep $\left(\ \tiny\yng(3,1,1,1,1)\ \right)$, whose $SU(3)_{c}\times SU(2)_{s}$ content is listed in Table~\ref{table2} line-4.
Only the   color triplets interest us.  The $[6]_{cs}$ contains a color triplet with spin-$1/2$ and the $[120]_{cs}$ contains color triplets with spins-$1/2$ and $3/2$.   

The wavefunctions of the spin-$3/2$ states are simple because only the spin-$1$ (color sextet) diquark, $\delta$, in the $[21]$ can couple to the antiquark to give spin-$3/2$,
\begin{equation}
\label{21spin3/2}
\left|[qq^{[21]}\bar q]\ [120](3_{c},4)\right> = \left|\delta \bar q, (3_{c},4)\right>
 \end{equation}
I have streamlined the notation as much as possible by omitting the flavor quantum numbers, which are understood to be $\mathbf{\overline 3}_{f}\otimes
\mathbf{\overline 3}_{f}=\mathbf{3}_{f}\oplus\mathbf{\overline{6}}_{f}$, by using the compact notation for diquarks defined in eq.~(\ref{quarkquark}), and by omitting any labels that are obvious from the context.  The superscript $[21]$ on the $qq$ label is necessary to distinguish these states from similar ones in the $[15]$ sector.

There are two spin-$1/2$ states in this multiplet, one in the $[6]$ and one in the $[120]$.  They are linear combinations of the $\alpha$ and $\delta$ diquark states.  It is convenient to use these diquark states as the basis states,
\begin{equation}
\begin{array}{lll}
\left|[qq^{[21]}\bar q]\ [6](3_{c},2)\right> &=& \sqrt{\frac{6}{7}}\left|\delta \bar q,  (3_{c},2)\right>
-\sqrt{\frac{1}{7}} \left|\alpha \bar q  ,  (3_{c},2)\right>\nonumber\\
\left|[qq^{[21]}\bar q]\ [120](3_{c},2)\right> &=& \sqrt{\frac{1}{7}}\left|\delta \bar q,   (3_{c},2)\right>
+\sqrt{\frac{6}{7}} \left|\alpha \bar q  ,   (3_{c},2)\right> 
\end{array}
\end{equation}
The colormagnetic interaction Hamiltonian can be evaluated in these basis states using eq.~(\ref{cscasimir}).  It mixes the $[6]$ and $[120]$ in the spin-1/2 sector.  The mixing and resulting eigenvalues and eigenstates were first calculated in Ref.~\cite{Mulders:1978cp} and are summarized in Table~\ref{tableqqqbar}.  The resulting eigenvectors can be written in terms of $\alpha$ and $\delta$-diquarks,
\begin{equation}
\begin{array}{lll}
\left|[qq^{[21]}\bar q] \Delta=-5.42\ (3_{c},2) \right> &=&  -0.582\left|\alpha \bar q, (3_{c},2) \right>
+0.813\left|\delta \bar q  ,  (3_{c},2) \right>\nonumber\\
\left|[qq^{[21]}\bar q] \Delta=-0.25\ (3_{c},2) \right> &=& \phantom{+}0.813\left|\alpha \bar q,  (3_{c},2) \right>
+0.582\left|\delta \bar q  ,  (3_{c},2) \right>
\end{array}
\label{21states}
\end{equation}
One of the spin-1/2 multiplets is anomalously light\footnote{The lightest $J=1/2$ $[qq\bar q]^{3_{c}}$ state is similar but not identical to the ``triquark'' discussed by Karliner and Lipkin.  Their state is pure $|\delta\bar q\>$ in our notation\cite{kl}{.  For a more general analysis of triquarks, see \cite{pentas}.}}; the other and the spin-3/2 multiplet are heavier.  

\begin{table}
\caption{$[qq\bar q]^{3_{c}}$ states in the  colorspin basis. $\Delta_{ij}$ is a matrix when two total colorspin irreps are mixed by colormagnetic interactions.}
\begin{tabular}{|c|c|c|c|c|}
\hline
 $SU(6)_{cs}$ Irrep &
      Spin & $SU(3)_{f}$ Irrep & $ {\Delta}_{ij}$& Eigenvalue(s) \\
\hline
[120]& $\frac{3}{2}$ & ${\mathbf{3}_{f}}\oplus\overline \mathbf{6}_{f}$ & $\frac{4}{3}{\cal M}_{qq\bar q}$&
$\frac{4}{3}{\cal M}_{qq\bar q}$\\
\hline 
\parbox[c]{1in}{\vskip 11pt\begin{center}$[6]$\\ $[120]$\end{center}}    
 & $\frac{1}{2}$ &  ${\mathbf{3}_{f}}\oplus\overline \mathbf{6}_{f}$ & $
 \left(\begin{array}{cc}-\frac{36}{7}  & \frac{10}{21}\sqrt{6}  \\ \frac{10}{21}\sqrt{6} & -\frac{11}{21}  \end{array}\right){\cal M}_{qq\bar q}$& 
\parbox{1in}{\begin{center}$ -5.42{\cal M}_{qq\bar q}$\\ $-0.25{\cal M}_{qq\bar q}$\end{center}} 
\\
\hline
$[84]$& $\frac{3}{2}$ & ${\mathbf{3}_{f}\oplus\mathbf{15}_{f}}$ & $\frac{4}{3}{\cal M}_{qq\bar q}$&
$\frac{4}{3}{\cal M}_{qq\bar q}$\\
\hline
 \parbox{1in}{\begin{center}$[6]$\\$ [84] $\end{center}}& $\frac{1}{2}$ &  ${\mathbf{3}_{f}}\oplus  \mathbf{15}_{f}$ & 
$\left(\begin{array}{cc} -\frac{12}{5}&\frac{2}{15}\sqrt{6}\\
\frac{2}{15}\sqrt{6} & \frac{41}{15}\end{array}\right){\cal M}_{qq\bar q}$
& 
\parbox{1in}{\begin{center} $-2.421{\cal M}_{qq\bar q}$\\$2.754{\cal M}_{qq\bar q}$\end{center}} 
\\
\hline 
\end{tabular}
\label{tableqqqbar}
\end{table}%
\begin{table}
\caption{$[qq\bar q]^{3_{c}}$ States in the schematic diquark model.}
\begin{tabular}{|c|c|c|c|}
\hline
Diquark content& Spin& $SU(3)_{f}$ Irrep& ``Mass''\\
\hline
$\alpha \bar q$ & $\frac{1}{2}$& $\mathbf{3}_{f}\oplus \overline\mathbf{6}_{f}$& $M_{qq\bar q}$\\
\hline
$\beta \bar q$ & $\frac{1}{2}\oplus\frac{3}{2}$& ${\mathbf{3}_{f}\oplus\mathbf{15}_{f}}$& $M_{qq\bar q}+
\Delta{\cal M}$\\
\hline 
\end{tabular}
\label{tableqqqbarschematic}
\end{table}%

The schematic diquark model ignores the $\delta$-diquark in the $[21]$ and therefore gives only a single spin-1/2 multiplet instead of two spin-1/2 and one spin-3/2.  However, looking at Table~\ref{tableqqqbar} we see that in the colormagnetic model one spin-1/2 multiplet is much more strongly bound ($\Delta=-5.42$) than the other ($\Delta=-0.25$) or the spin-3/2 multiplet ($\Delta=+4/3$).  Since only the lightest multiplet is likely to be stable enough to find in a lattice simulation, the two pictures effectively agree on the spectrum of prominent states.  However there is a striking difference in diquark content:  In the schematic model the lightest state is pure $\alpha$-diquark by hypothesis.  In the colormagnetic model, according to eq.~(\ref{21states}) the light state includes a significant admixture (approximately $2:1$ in probability) of the color sextet $\delta$-diquark.  Apparently the color magnetic interaction between the quarks and the antiquark overwhelm the diquark correlation energy.  Even though the spectroscopic consequences are the same, the internal correlations differ dramatically between the two models.  The same pattern persists in the colorspin antisymmetric ($[15]$) diquark sector.

\bigskip
2.\quad  $([15]\ {\mathbf{6}}_{f})\otimes ([\overline 6]\ {\mathbf{\overline{3}}_{f}})$
\medskip

These states are in the $ {\mathbf{6}}_{f} \otimes {\mathbf{3}}_{f} \ = \ \mathbf{3}_{f}\oplus {\mathbf{15}}_{f}$ of flavor.
The Clebsch Gordan series for $[15]\times[\overline 6]$ includes the fundamental and an 84-dimensional irrep $\left(\ \tiny\yng(2,2,1,1,1)\ \right)$, whose $SU(3)_{c}\times SU(2)_{s}$ content is listed in Table~\ref{table2} line-5.
Only the color triplets interest us.  As in the previous case there is a spin-3/2 state in the large irrep, while spin-1/2 occurs in both. 
The wavefunction of the spin $3/2$ state  is simple:
\begin{equation}
\label{18spin3/2}
\left|[qq^{[15]}\bar q]\ [84](3_{c},4) \right> = \left|\gamma \bar q  ,  (3_{c},4)\right>
 \end{equation}
As before, flavor quantum numbers ( $\mathbf{3}_{f}\oplus\overline{\mathbf{15}}_{f}$) have been suppressed.

The  spin-$1/2$ states  are linear combinations of the $\beta$ and $\gamma$ diquark states.  Using these diquark states as the basis states,
\begin{equation}
\begin{array}{lll}
\left|[qq^{[15]}\bar q] [6](3_{c},2) \right> &=& \sqrt{\frac{2}{5}}\left|\gamma \bar q, (3_{c},2) \right>
-\sqrt{\frac{3}{5}} \left|\beta \bar q  ,  (3_{c},2) \right>\nonumber\\
\left|[qq^{[15]}\bar q] [84](3_{c},2) \right> &=& \sqrt{\frac{3}{5}}\left|\gamma \bar q,  (3_{c},2) \right>
+\sqrt{\frac{2}{5}} \left|\beta \bar q  ,  (3_{c},2) \right>\label{15mixing}
\end{array}
\end{equation}
Color magnetism favors one of the spin-1/2 multiplets (see Table~\ref{tableqqqbar}).  The two spin-1/2 eigenstates are both linear superpositions of $|\beta\bar q\>$ and $|\gamma\bar q\>$:
\begin{equation}
\begin{array}{lll}
\left|[qq^{[15]}\bar q] \Delta=-2.421\ (3_{c},2) \right> &=&  -0.582\left|\gamma \bar q, (3_{c},2) \right>
+0.813\left|\beta \bar q  ,  (3_{c},2) \right>\nonumber\\
\left|[qq^{[15]}\bar q] \Delta=2.754\ (3_{c},2) \right> &=& \phantom{+}0.813\left|\gamma \bar q,  (3_{c},2) \right>
+0.582\left|\beta \bar q  ,  (3_{c},2) \right>
\label{15states}
\end{array}
\end{equation}
Again the lightest state has spin-$1/2$, but it is considerably heavier than the lightest spin-$1/2$ multiplet built of $\alpha$ and $\delta$ diquarks (see eq.~({\ref{21states})).  
\begin{floatingfigure}{.65\textwidth} 
\includegraphics[width=9cm]{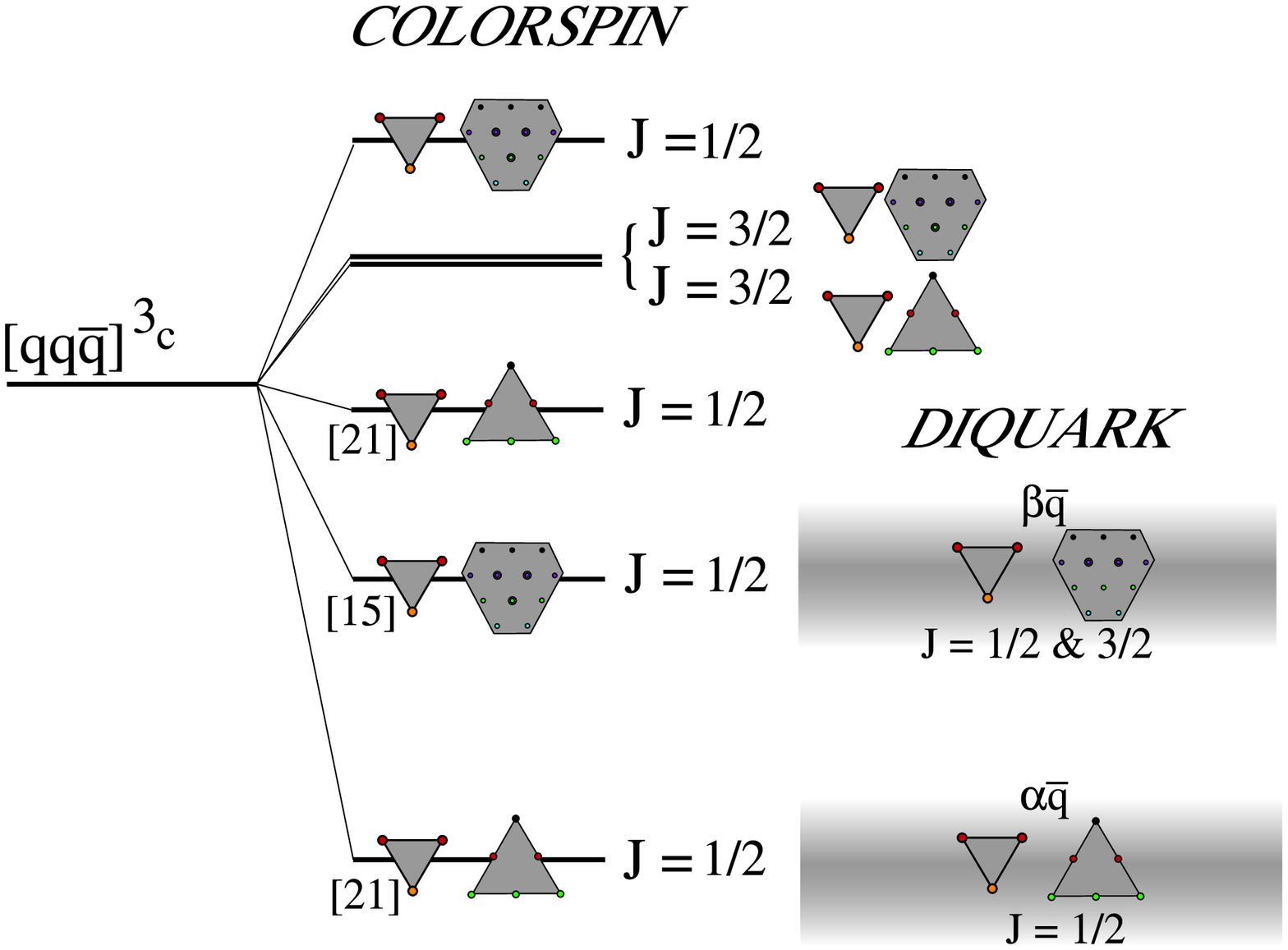}
\caption{\small  The spectrum of $[qq\bar q]^{{3}_{c}}$ states.  The first column shows colormagnetic model splittings.  The second gives a sketch of the ordering of states in the schematic diquark model.  $SU(3)_{f}$ multiplets are labeled by icons of their weight diagrams.}
  \label{qqqbar}
\end{floatingfigure}
The schematic diquark model ignores the $\gamma$ diquark entirely and therefore yields only one spin-1/2 multiplet and one spin-$3/2$ multiplet.  Since the $\beta$-diquark is assumed heavier than the $\alpha$-diquark, this multiplet is expected to be heavier than the $[\alpha\bar q]$ state found in the previous subsection.  Once again the schematic diquark model mimics the colormagnetic model at least as far as the light states are concerned.  As before the distinction between the models lies in the diquark content of the light state:  in the schematic model it is pure $\beta$-diquark, while colormagnetism intermixes the color-sextet $\gamma$-diquark approximately $1:2$ relative to the $\beta$-diquark\footnote{The coefficients in eqs.~(\ref{21states}) and (\ref{15states}) are in fact identical.  They are numerically close but not equal to $\sqrt{1/3}$ and $\sqrt{2/3}$.}
 
The spectrum of $[qq]\bar q$ states in the two models is summarized in Fig.~\ref{qqqbar}.  In both models the lightest multiplet is a degenerate $\mathbf{3}_{f}$ and $\mathbf{\overline{6}}_{f}$ with spin-$1/2$.  The spectra begin to differ at the next level, although even there the differences are minor. 

\section{Discussion and Conclusions}

There is little doubt that QCD is the correct theory of hadrons.  Nor is there much doubt  that where lattice simulations can be applied, they accurately describe QCD.  The quality of lattice simulations of QCD has been increasing for years, so one might expect that the days of phenomenological models are over.  In the not too distant future (if not now) experiment will be compared directly with lattice calculations.  However  phenomenological models will not be so quickly dispatched:  They provide  \emph{physical} insight into the  complex phenomena associated with confinement that has not been possible to extract from lattice simulations alone.  This is particularly clear for light ($u$, $d$, $s$) quarks.  Chiral dynamics, constituent quark and large $N_{c}$ spectroscopic regularities, Regge phenomenology, quark counting rules, QCD sum rules, vector dominance, {\it etc.\/}, carry our understanding of QCD beyond the regime where lattice calculations have yet been applied.  

If there is a place for phenomenology in a modern discussion of QCD, then we need to know how to improve our phenomenological models.  They were first developed in the 1960's and 1970's in response to a flood of experimental data, but qualitatively new data is relatively rare now.
We need new ``data'' to test and refine them.  It is natural, then, to suggest that lattice simulations be used to explore novel regimes or parametrically different versions of QCD which are not accessible to experiment, but which can test the validity of phenomenological models.  Non-singlet spectroscopy is a case in point.  It can be formulated on the lattice with the help of color neutralizing Wilson lines.  Phenomenological models of light quark dynamics can take a crack at estimating the spectrum and static properties of the light states.  There are probably numerous other examples of problems where something can be learned by comparing models and lattice simulations in regimes where experiments are presently (or permanently) impossible.

In this paper I have mapped out the terrain in two simple color triplet sectors of QCD spectroscopy:  $[qqqq]^{3_{c}}$ and $[qq\bar q]^{3_{c}}$.  Both have a rich spectrum and offer the possibility to contrast different pictures of $qq$-correlations in QCD.  If the light states in these sectors are relatively stable, then their phenomenology might be nearly as instructive as the spectroscopy of the light mesons and baryons.  They can be studied on the lattice by constructing color triplet sources in the presence of a neutralizing Wilson line.  Since the Wilson line has to begin and end on a definite lattice site, it is more difficult to accumulate statistics for such a calculation than for a more conventional correlator.  Newly developed ``all-to-all'' propagators offer hope of improving this situation\cite{Foley:2005ac}.

If they are too massive, color triplet states can fall apart into mesons or baryons plus a single quark:  $[qq\bar q]^{3_{c}}\to [q\bar q] q^{3_{c}}$ or $[qqqq]^{3_{c}}\to [qqq] q^{3_{c}}$.  Indeed, the failure, so far, to find experimental evidence for heavy quark exotics with quark content $qqqq\overline Q$ or $qq\bar q\overline Q$ suggests that none of these states are \emph{very} stable for physical values of the quark masses.  However states with typical hadronic widths are not excluded and furthermore, lattice simulations can change quark masses to values where stability may be more favorable.  Even one or two states stable enough to study on the lattice would provide important information for the iimprovement of phenomenological models.

\section{Acknowledgments}
I would like to thank D.~Boer, K.~J.~Juge, L.~Maiani, J.~Negele, M.~Peardon, F. Piccinini, A.~D.~Polosa, and F.~Wilczek  for conversations and encouragment.  I am particularly grateful to Daniel Boer for reminding me of Ref.~\cite{Mulders:1978cp}.
This work is  
supported in part by funds provided by the U.S.~Department of
Energy (D.O.E.) under cooperative research agreement
DE-FC02-94ER40818 and in part by the INFN under the MIT/INFN Bruno Rossi Exchange Program.

\section{Appendix} 

Here is a simple way to compute the diquark content of $[qqq]$ and $[qqqq]$ states.  In the hadrons of interest to us all the quarks are equivalent.  Therefore
\begin{equation}
\<\vec\sigma_{1}\cdot\vec\sigma_{2}\> = \frac{2}{N(N-1)}\<\sum_{i>j}
\vec\sigma_{i}\cdot\vec\sigma_{j}\>=\frac{1}{N(N-1)}(4S(S+1)-3N)
\label{spin}
\end{equation}
where  $\vec\sigma_{1}$ and $\vec\sigma_{2}$ are (twice) the spins of any pair of quarks  and, of course, $S(S+1)$ is the quadratic Casimir of $SU(2)$.  The analagous formulas for the product of two quarks' color ($\tilde\beta_{1}\cdot\tilde\beta_{2}$) and color$\times$spin ($(\vec\sigma\tilde\beta)_{1}\cdot(\vec\sigma
\tilde\beta)_{2}$) are,
\begin{equation} 
\<\tilde\beta_{1}\cdot\tilde\beta_{2}\> = \frac{2}{N(N-1)}\<\sum_{i>j}
\tilde\beta_{i}\cdot\tilde\beta_{j}\>=\frac{4}{N(N-1)}\left({\cal C}_{c}-\frac{4}{3}N\right)
\label{color}
\end{equation}
and 
\begin{eqnarray} 
\<(\vec\sigma\tilde\beta)_{1}\cdot(\vec\sigma\tilde\beta)_{2}\> &=& \frac{2}{N(N-1)}\<\sum_{i>j}
(\vec\sigma\tilde\beta)_{i}\cdot(\vec\sigma\tilde\beta)_ {j}\>\nonumber\\
&=&\frac{4}{N(N-1)}\left({\cal C}_{cs}-
\frac{2}{3}S(S+1)-{\cal C}_{c}-4N\right)
\label{colorspin}
\end{eqnarray}
The values of these matrix elements for each of the four diquarks are given in Table~\ref{innerprods}.
\begin{table}
\caption{Useful diquark matrix elements.}
\begin{tabular}{|c|c|c|c|}
\hline
Diquark & $\<\vec \sigma_{1}\cdot\vec\sigma_{2}\>$& $\<\tilde\beta_{1}\cdot\tilde\beta_{2}\>$ &$\<(\vec\sigma\tilde\beta)_{1}\cdot(\vec\sigma
\tilde\beta)_{2}\>$  \\
\hline
$\alpha $ & $-3$& $-\frac{8}{3}$& $\phantom{-}8$\\
\hline
$\beta $ & $\phantom{-}1$& $-\frac{8}{3}$& $-\frac{8}{3}$\\
\hline 
$\gamma $ & $-3$& $\phantom{-}\frac{4}{3}$& $-4$\\
\hline
$\delta $ & $\phantom{-}1$& $\phantom{-}\frac{4}{3}$& $
\phantom{-}\frac{4}{3}$\\
\hline
\end{tabular}
\label{innerprods}
\end{table}%

When the matrix elements of eqs.~(\ref{spin}), (\ref{color}), and (\ref{colorspin}) are evaluated in any state composed of equivalent quarks, the results must be a weighted average of the probability that any pair of quarks find themselves in the $\alpha$, $\beta$, $\gamma$, and $\delta$ configurations.  The resulting equations can then be solved for these weights.

As a warm-up consider the color singlet states of $qqq$.  The multiplets made of three equivalent $j=1/2$ quarks are the familiar ground state octet (spin-$1/2$) and decuplet (spin-$3/2$) baryons.    It suffices to consider $\<\vec\sigma_{1}\cdot\vec\sigma_{2}\>$.  Using eq.~(\ref{spin}) we obtain
\begin{eqnarray}
\left.\<\vec\sigma_{1}\cdot\vec\sigma_{2}\>
\right|_{(1_{c},2)\ \mathbf{8}_{f}}&=&-1 \nonumber\\
\left.\<\vec\sigma_{1}\cdot\vec\sigma_{2}\>
\right|_{(1_{c},4)\ \mathbf{10}_{f}}&=&+1  
\label{qqq-content}
\end{eqnarray}
where we have labeled the $ qqq $ states using the (color,spin){\bf flavor} notation used for color triplet states in the paper.
Any given diquark in a color singlet $ qqq $ state must be a color antitriplet.  This eliminates the $\gamma$ and $\delta$-diquarks.  So the results of eq.~(\ref{qqq-content}) must be a weighted average of the $\alpha$ and $\beta$ diquarks,
\begin{eqnarray}
1&=& -3  P_{\alpha}((1_{c},4_{s})\ \mathbf{10}_{f})+   P_{\beta} ((1_{c},4_{s})\ \mathbf{10}_{f})\nonumber\\
-1&=& -3  P_{\alpha}((1_{c},2_{s})\ \mathbf{8}_{f} )+ P_{\beta} ((1_{c},2_{s})\ \mathbf{8}_{f})\nonumber\\
\label{qqqweighting}
\end{eqnarray}
where $P_{\alpha}(S)$ ($P_{\beta}(S)$) is the probability that two quarks form the $\alpha$ ($\beta$) diquark in the state $S$.
This gives the standard result that any diquark in the decuplet must be purely $\delta$ (which is the only way to make spin-$3/2$), and any diquark in the octet must, on average be 50\% $\alpha$ and 50\% $\delta$.

$[qqqq]^{{3}_{c}}$ is the non-trivial case.  Combining the diquark matrix elements in Table~\ref{innerprods} with eqs.~(\ref{spin}), (\ref{color}) and (\ref{colorspin}), we obtain a result for any color triplet state of $N$-equivalent quarks,
\begin{eqnarray}
\label{master}
P_{\alpha}+P_{\beta}+P_{\gamma}+P_{\delta} &=& 1\nonumber\\
-3P_{\alpha}+P_{\beta}-3P_{\gamma}+P_{\delta} &=& 
\frac{1}{3}S(S+1)-1\nonumber\\
-2P_{\alpha}-2P_{\beta}+P_{\gamma}+P_{\delta} &=& -1\nonumber\\
8P_{\alpha}-\frac{8}{3}P_{\beta}-4P_{\gamma}+\frac{4}{3}P_{\delta} &=& \frac{1}{3}\left({\cal C}_{cs}-\frac{2}{3}S(S+1)-\frac{52}{3}\right)\nonumber\\
\end{eqnarray}
Substituting the appropriate Casimirs for the $[qqqq]^{{3}_{c}}$ representations in Table~\ref{qqqq}, we obtain the probabilities quoted in the right most columns of that Table.

%
%


\begin{thebibliography}{99}
\bibitem{jwn} I thank J.~Negele for pointing out this alternative and for discussions on implementing it.
\end{thebibliography}
  \end{document}